\documentclass[aps,floatfix,showpacs,prl,twocolumn]{revtex4}
\usepackage{graphicx,bm}

\begin{document}

\title{Quantum correlations of two optical fields close to  
       electromagnetically induced transparency}  

\author{A. Sinatra}
\affiliation{ \small Laboratoire Kastler Brossel, ENS,
 24 Rue Lhomond, 75231 Paris Cedex 05, France}

\begin{abstract}
We show that three-level atoms excited by two cavity modes in a $\Lambda$
configuration close to electromagnetically induced transparency can
produce strongly squeezed bright beams or correlated beams which can
be used for quantum non demolition measurements.
The input intensity is the experimental ``knob" for tuning the system into
a squeezer or a quantum non demolition device. The quantum correlations
become ideal at a critical point characterized by the appearance of a 
switching behavior in the mean fields intensities.
Our predictions, based on a realistic fully quantum 3-level model 
including cavity losses and spontaneous emission, allow direct comparison 
with future experiments.
\end{abstract}


\maketitle
Using optical nonlinearities it is possible to manipulate optical beams
to the level of quantum fluctuations, producing squeezed 
states \cite{squeezing} that are important resources for quantum information 
with continuous variables \cite{contvar}. 
Related to the field of squeezing is that of quantum non
demolition (QND) measurements on an optical field, where quantum
correlations between two different modes of the electromagnetic field
are exploited to overcome the back-action noise of a quantum measurement
\cite{QND}. Besides the fundamental interest in the theory of measurement, 
it was shown that QND correlations of propagating beams have direct 
application in quantum communication protocols as teleportation \cite{Kilin}.
The best single back-action-evading measurement on optical beams was
performed using cold atoms inside a doubly resonant cavity \cite{PRL_QND}.
We suggest that these performances could be significantly improved 
by tuning the system close to the electromagnetically
induced transparency (EIT) conditions \cite{EIT}.

\begin{figure}[htb]
\centerline{\includegraphics[width=6cm,clip=]{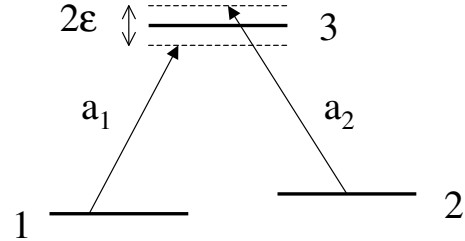}}
\caption{Two cavity modes interact with the atoms in a $\Lambda$ configuration close to EIT conditions.}
\label{fig:Lambda}
\end{figure}

Already in the nineties, theoretical studies 
showed that a lambda three-level medium close to EIT conditions
in a cavity can be used to obtain squeezing \cite{Gheri-Marte}. 
Contrarily to previous proposals, here we assume that two different modes are
resonant in the cavity. For small and symmetrical detunings from the upper
level of $\Lambda$ three-level atoms (see Fig.\ref{fig:Lambda}), absorption
is suppressed and the dispersive non linear response 
gives rise to a rich scenario
where either self correlations (squeezing) or cross QND correlations
can be established in the output beams. The correlations become ideal 
at a critical point that we characterize analytically.
The technique we propose is experimentally accessible, and first experimental
steps in this directions were done in atomic vapors without a cavity \cite{Scully}.
Here we show that the presence of the cavity is a crucial advantage especially
if one can reach the good cavity limit.

We consider $N$ atoms in a cavity excited by two modes symmetrically 
detuned from the upper level of a $\Lambda$ three-level scheme as in Fig.\ref{fig:Lambda}.
For $j=1,2$ let $\omega_j$ be the frequency of field $j$ and 
$\hbar \omega_{aj}$ the energy of the corresponding atomic transition. 
We define $\Delta_j = \frac{\omega_{aj}-\omega_{j}}{\gamma_w}$ the atomic detunings
normalized to the decay rate of the optical coherences $\gamma_w=(\gamma_1+\gamma_2)/2$
where $\gamma_1+\gamma_2$ is the total population decay rate of the upper level; 
$\theta_j = \frac{\omega_{cj}-\omega_{j}}{\kappa_j}$ the cavity detunings   
normalized to the cavity decay rates $\kappa_j$, and  
$C_j = \frac{ g_{j}^{2} N}{\gamma_w \kappa_j}$ the cooperativities where $g_j$ 
are the coupling constants for the two considered transitions.
We use normalized variables proportional to the intracavity and input fields
$x_{j}  =  \frac{\sqrt{2}g_{j}}{\gamma_w}\langle a_{j} \rangle$ and 
$y_{j}  =  \frac{\sqrt{2}g_{j}}{\gamma_w} \frac{2}{\sqrt{T_j}}E_j^{in}$ respectively,
where $T_j$ is the (input-output) mirror transmissivity for the field $j$.
We name $v$ and $w$ the normalized polarizations between levels 1-3 and 2-3
$v=-(\sqrt{2}/N)\langle R^- \rangle$, $w=-(\sqrt{2}/N)\langle S^- \rangle$
where $R$ and $S$  are collective operators constructed from the single
atom operators $|1\rangle \langle 3|$ and 
$|2\rangle \langle 3|$ as in \cite{ProgrOpt}.
The master equation and the semiclassical equations describing the $\Lambda$ system with 
two cavity fields, with the same notations introduced here,
are given and discussed in detail in \cite{PRA_QND} 
where this model was successful to reproduce the experimental results of \cite{PRL_QND}.

\begin{figure}
\centerline{\includegraphics[width=6cm,clip=]{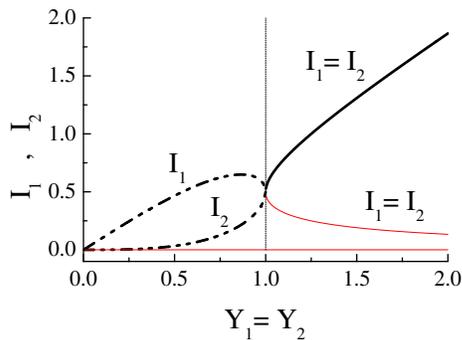}}
\caption{Stationary intensities of the intracavity fields $I_1$, and $I_2$
as a function of the common intensity of the input fields $Y_1=Y_2=Y$. In solid line
the solution $I_1=I_2$. The thick (red thin) line correspond to stable 
(unstable) solutions. In dashed-dotted line one of the two stable
solutions with $I_1 \neq I_2$. 
Parameters: $2\epsilon=0.125$ and $C_1=C_2=250$, 
$\gamma_1=\gamma_2=10 \kappa_1$, $\kappa_2=\kappa_1$,
$\theta_1=\theta_2=0$.}
\label{fig:bista}
\end{figure}

Let us consider a set of parameters symmetric for the two transitions: 
$|y_j|=|y|$, $C_j=C$, $\gamma_j=\gamma$, $\kappa_j=\kappa$, $\theta_j=0$ (empty cavity
resonance for both fields), and let $\Delta_1=-\Delta_2=\epsilon$ be small and positive.
In Fig. \ref{fig:bista} we show in rescaled units
the stationary intensities of the intracavity fields $I_j=|x_j|^2/4C\epsilon$ as   
a function of the common intensity of the input fields $Y=|y|^2/4C\epsilon$. 
With a solid line we have plotted an $S$-shaped solution with $I_1=I_2$. 
A stable branch of this solution appears for $Y>1$. 
The negative slope branch and the lower branch very close to zero intensity 
are both instable and play no role in the following. 
For $Y<1$, apart from the solution $I_1=I_2$, we get two other
solutions with $I_1\neq I_2$. 
In the figure we show one of them with $I_1>I_2$. The second one is obtained by
exchanging $I_1$ and $I_2$. Both solutions 
are stable in the considered case $\theta_1=\theta_2=0$.

\begin{figure}[htb] 
\centerline{\includegraphics[width=8cm,clip=]{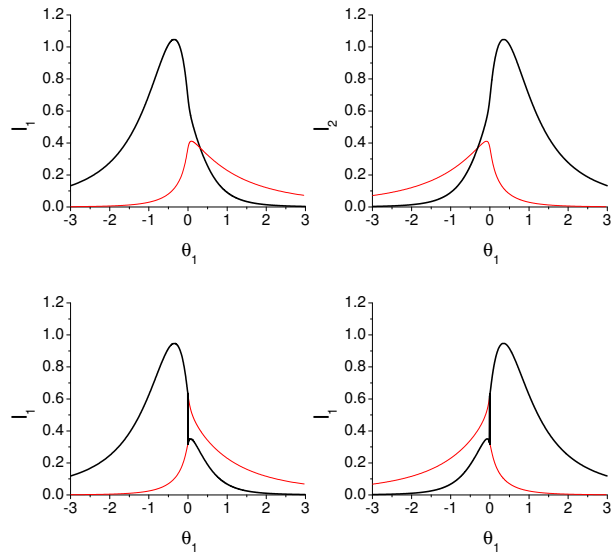}}
\caption{Intracavity field intensities $I_1$ (left half), $I_2$ (right half)
across the cavity scan. Upper half: $Y=1.05$.
Lower half: $Y=0.95$. The thick (red thin) lines correspond to stable (unstable)
solutions. The other parameters are as in Fig.~\ref{fig:bista}.}
\label{fig:switch}
\end{figure}

We choose now two values of the input intensity,  
in turn above and below the turning point  $Y=1$,
and show the stationary solutions for intracavity fields
intensities as the cavity detunings vary in Fig.~\ref{fig:switch}. 
The stable branches of these curves (thick lines)
can be easily obtained experimentally by sweeping the cavity length 
\cite{PRA_QND}. We vary $\theta_1$ and  $\theta_2$ keeping them always equal 
which would imply the use of two driving fields of close optical frequencies 
$\Delta\lambda/\lambda \ll 1$ (and for example different polarizations).
For $Y=1.05$ i.e. 5\% {\it above} the turning point 
(upper half of Fig.~\ref{fig:switch})
the stable solutions for the intracavity intensities are  
Lorentzian-looking curves symmetrically shifted by a small amount from their 
empty-cavity positions for both fields. Only for $\theta_1=\theta_2=0$ 
the two fields have the same stationary amplitude in the cavity
corresponding to the stable high-transmission branch
of the $S$-shaped curve in Fig.\ref{fig:bista}.
For $Y=0.95$ i.e. 5\% {\it below} the turning point 
(lower half of Fig.~\ref{fig:switch})
the situation is rather different: 
the stable solution for the two fields {\it switches} between a high-intensity
and a low-intensity curve being always $I_1 \neq I_2$ although $|y_1|=|y_2|$.
In contrast with the previous case this situation is very far
from the independent-fields EIT solution and the fields are in fact
strongly coupled. 

Let us now introduce the usefull correlations to 
caracterize the quantum fluctuation properties of the system.
For a given quadrature of the $j^{th}$ 
field: $X_j^\phi~=~a_j e^{-i\phi} + i a^{\dagger}_j e^{i\phi}$,
the squeezing spectrum is defined as
\begin{equation}
S^{\phi}_{j}(\omega)=1+2\kappa_j \int_{-\infty}^{\infty} e^{-i\omega t}
\langle \, :\delta X_{j}^\phi (t) \, \delta X_{j}^\phi(0): \, \rangle  \, dt
\label{eq:spettro}
\end{equation}
where $\delta X_j^\phi$ denotes the time dependent fluctuation
of the operator $X_j^\phi$ around a steady state point.
The column indicates normal and time ordering for the product
inside the mean. $S_j^\phi=1$ is the shot noise and $S_j^\phi=0$ means total suppression 
of fluctuations in the quadrature $X_j^\phi$.
The crossed correlations between the two fields are described by 
the coefficients $C_s$, $C_m$ and $V_{s|m}$ \cite{QNDCoeff} 
characterizing a QND measure of the amplitude quadrature $X^{in}$ of one field,
the {\it signal}, performing
a direct measurement on the phase quadrature $Y^{out}$ of the other
field, the {\it meter}.
Among the three coefficients 
$C_s$ quantifies the non-destructive character of the measurement,
$C_m$ its accuracy and $V_{s|m}$ refers to the  to the 
``quantum state preparation" capabilities of the system.
\begin{eqnarray}
C_s &=&  C(X^{in},X^{out})  \,,
\hspace{0.5cm} 
C_m =  C(X^{in},Y^{out})\,,  \\
V_{s|m} &=&  \langle X^{out},X^{out} \rangle
\left(1- C(X^{out},Y^{out}) \right)  
\end{eqnarray}
where for two operators $A$ and $B$ we define
\begin{equation}
C(A,B)=\frac{|\langle A,B \rangle |^2}{\langle A,A \rangle  \langle B,B \rangle } \hspace{0.5cm}
\mbox{with}
\end{equation}
\begin{equation}
\langle A,B\rangle=\int_{-\infty}^{+\infty} e^{-i\omega t}
\frac{1}{2}\langle A(t) B+ B A(t)\rangle\, dt \,. 
\end{equation}
Superscripts $in$ and $out$ refer to the input and output cavity fields.
By calling $\phi_j^{in}$ and $\phi_j^{out}$ the phases of the input and output fields in 
steady state, and choosing field 1 as the {\it meter} and field 2 as the {\it signal},
we define $X^{out(in)}=X_2^{\phi_2^{out(in)}}$ and
$Y^{out}=Y_1^{\phi_1^{out}+\pi/2}$.
For an ideal QND measurement $C_m=C_s=1$, and $V_{s|m}=0$.

\begin{figure}[htb]
\centerline{\includegraphics[width=5.5cm,clip=]{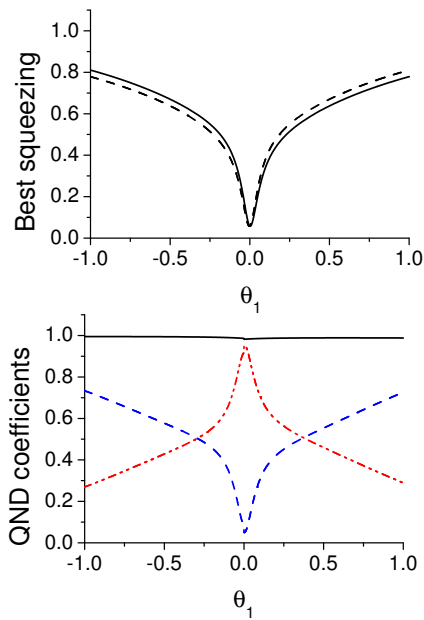}}
\caption{
Top: Best squeezing of the fields across the cavity scan 
for $Y=1.05$ and $\omega=0$. 
Squeezing of field 1 (2) is plotted with a solid (dashed) line. 
Bottom: QND coefficients across the cavity scan 
for $Y=0.95$ and $\omega=0$. 
$C_m$ (red dashed-dotted line), $C_s$ (solid line),  $V_{s|m}$ 
(blue dashed line). 
Parameters as in Fig.\ref{fig:switch}.}
\label{fig:quant}
\end{figure}

The quantum fluctuations counterpart of Fig.~\ref{fig:switch} (top)
is shown in Fig.~\ref{fig:quant} (top) where squeezing of the output fields 
optimized with respect to the quadrature $S_j^{best}(\omega=0)$ is plotted
as a function of the cavity detuning. 
A large amount of squeezing is present in both fields close to $\theta_1=0$.
As one can see from Fig.~\ref{fig:switch} (top) the two fields are well transmitted 
by the cavity for $\theta_1=0$, and the system efficiently converts the input 
coherent beams into bright squeezed beams. 
Correspondingly to Fig.~\ref{fig:switch} (bottom) for $Y=0.95$,
in Fig.\ref{fig:quant} (bottom) we plot the coefficients $C_s$, $C_m$ and $V_{s|m}$ 
across the cavity scan. The useful quantum correlations are calculated by 
a linearized treatment of quantum fluctuations
around the stable stationary solution as in \cite{PRA_QND}.
Despite the fact that the two fields have different 
intracavity intensities at $\theta_1=0$,
they play here symmetrical roles for the QND scheme;
the figure corresponding to the reversed scheme $1\leftrightarrow 2$ being 
obtained by reflection of the plots $\theta_1 \leftrightarrow - \theta_1$. 

\begin{figure}[htb]
\centerline{\includegraphics[width=5.5cm,clip=]{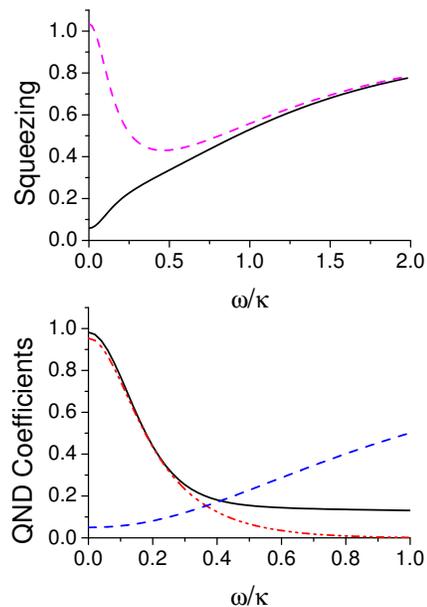}}
\caption{
Top: Squeezing spectra of field 1, for $Y=1.05$ 
in the center of the cavity scan ($\theta_1=0.0013$).
Best squeezing in solid line and 
amplitude squeezing $S_1^{\phi_1^{out}}$ in purple dashed line. 
Bottom: QND spectra for $Y=0.95$ and $\theta_1=0.0018$. 
$C_m$ (red dashed-dotted line), $C_s$ (solid line),  $V_{s|m}$
(blue dashed line).
The other parameters are as in Fig.\ref{fig:bista}.}
\label{fig:spectra}
\end{figure}

We show in Fig.~\ref{fig:spectra} the frequency dependence of the 
quantum correlations both below and above the turning point $Y=1$,
for a fixed value of the cavity detuning close to zero. 
For values of the cooperativity parameters currently obtained in experiments, 
QND coefficients such as $C_s=0.98$, $C_m=0.95$, $V_{s|m}=0.05$
can be acheived in this regime, representing a significant improvement
with respect to previously obtained results \cite{PRL_QND}
based on the so called ``ghost transition" scheme \cite{Gheri}, \cite{PRA_QND}.
Although we concentrate here on the good cavity limit, in which as we will show
the quantum correlations become ideal approaching the turining point $Y=1$,
some QND correlations between the two modes persist also in the
bad cavity limit. For example for $\epsilon=0.25$, $C=25$, $\kappa=3 \gamma$,
$\theta_1=6 \times 10^{-3}$, $Y=0.9$ and $\omega=0.1 \gamma$ we get
$C_s=C_m=0.72$, $V_{s|m}=0.26$. 

In the limit of weak atomic detunings, useful analytical
results can be obtained.
The analytical solution of the semiclassical equations of the system 
at steady state is given in \cite{PRA_QND}.
By expanding the steady state polarizations $v$ and $w$ 
to the first order in $\epsilon$ we obtain 
\begin{equation}
v= i \frac{ 4\epsilon \, x_1 |x_2|^2}{(|x_1|^2+|x_2|^2)^2}  \hspace{0.5cm}
w= - i \frac{ 4\epsilon \, x_2 |x_1|^2}{(|x_1|^2+|x_2|^2)^2} \,.
\label{eq:v_w}
\end{equation}
By inserting (\ref{eq:v_w}) in the equations 
for the intracavity fields 
amplitudes, with $|y_j|=|y|$, $\theta_j=0$ and $C_j=C$,
we obtain at steady state a ``universal solution" for rescaled field intensities.
For $Y<1$ there are two stable solutions 
\begin{eqnarray}
I_1 &=&\frac{Y}{2} \left( 1 \pm \eta \right)\,; \hspace{0.3cm} 
I_2 =\frac{Y}{2} \left( 1 \mp \eta \right) \label{eq:sol1}  
\end{eqnarray}
where $\eta=\sqrt{1-Y^2}$. For $Y>1$, out of two solutions 
\begin{equation}
I_1=I_2=I \:\: ;  \:\:
I=\frac{Y}{2} \left( 1 \pm \sqrt{1-\frac{1}{Y^2}} \right)
\label{eq:I}
\end{equation}
the one with the plus sign is stable and the other one unstable.
Solutions (\ref{eq:sol1})-(\ref{eq:I}) are indistinguishable from 
those of the full three-level model in Fig.~\ref{fig:bista}.
The phases of the input fields with respect to intracavity fields
(which are taken real at steady state) are
$\phi_1^{in}=\mbox{atan}\left[\sqrt{I_2/I_1}\right]$,  
$\phi_2^{in}=-\mbox{atan}\left[\sqrt{I_1/I_2}\right]$ for $Y<1$, and
$\phi_1^{in}=\mbox{atan}\left[ 1/2I \right]=-\phi_2^{in}$
for $Y>1$. For the output fields 
$\phi_1^{out}=-\phi_1^{in}$, $\phi_2^{out}=-\phi_2^{in}$
in both cases.

In order to study the quantum properties of the system analytically
we further assume that (i) $\gamma \gg \kappa$ so that the
atomic fluctuations follow adiabatically the field fluctuations,
and (ii) the noise from spontaneous emission is negligible,
which we found true when the cooperativity is large enough.
In this limit, using the steady state polarizations (\ref{eq:v_w}),
we can solve analytically the equations for the
field fluctuations and obtain the correlation functions. 

For  $Y>1$ and $I_1=I_2=I$ and taking $\kappa^{-1}$ as the unit of 
time, we obtain
\begin{equation}
\delta \dot{x}_j = - \delta x_j + i \, \frac{(-1)^{3-j}}{2 I} \delta x_j^{\ast} \,
\hspace{1cm} j=1,2 \,.
\label{eq:1flymag1} 
\end{equation}
These equations describe two independent two-photon processes for which
instabilities and squeezing have been studied extensively
\cite{Nard}. The best squeezing spectrum for each field is
\begin{equation}
S_j^{best}(\omega)=1-\frac{4a}{(1+a)^2+{\omega/\kappa}^2}\;,
\hspace{1cm} a=\frac{1}{2I}\;, 
\end{equation}
yielding perfect squeezing at zero frequency at the turning 
point where $Y=1$, $I=0.5$ and $a=1$. 

For $Y<1$ and  $I_1\neq I_2$ the fluctuations of the two fields are coupled.
For $I_1>I_2$ we get 
\begin{eqnarray}
\delta X_1 &=& - \delta X_1 - i \, \frac{1-\eta}{Y} \, \delta Y_1 \label{eq:deltax1}\\
\delta Y_1 &=& -\delta Y_1 + i  \frac{1+\eta}{Y} \, \delta X_1 - 2 i \, \eta \delta X_2 \,.
\label{eq:deltay1}
\end{eqnarray}
The equations for field 2 are obtained from (\ref{eq:deltax1}) and (\ref{eq:deltay1})  
by changing the sign in front of $\eta$ and of $i$.
Simple analytical expressions can be obtained for the 
squeezing and the conditional variance $V_{s|m}$
of the fields at $\omega=0$
\begin{eqnarray}
S_j^{int}&=&S_j^{best}=1 \:\: ; \:\: S_j^{phase}=-3+\frac{4}{\eta^2} \\
V_{s|m}&=&\frac{\eta^2}{4-3\eta^2}
\end{eqnarray}
showing that the fields have diverging phase noise and  
become perfectly correlated at the turning point.
We checked that the spectra in Fig. \ref{fig:spectra} are well reproduced by 
the analytical results.
 
In conclusion, in a symmetrically detuned EIT scheme, and for equal input
intensities $Y$ of the two fields
we have shown the existence of a universal $S$-shaped steady state curve (Fig.\ref{fig:bista})
which divides the parameter space into two parts:
for input intensities higher than the upper turning point of the curve,
the quantum fluctuations of the fields become quadrature dependent and can be
reduced in a quadrature, while for input intensities lower than the turning point,
crossed phase-intensity quantum correlations build up between the two fields.
The system becomes a perfect ``squeezer" or an ideal QND device at the turning point.
The ``universal" point $Y=1$, can be
identified experimentally by the appearance of the switching behavior
described in Fig.\ref{fig:switch}, and can be used as a reference
in the parameter space to choose either the squeezing or the QND effect
and to optimize it.
An implementation using either a vapor \cite{Scully},
or a trapped cold atoms in an optical cavity
\cite{PRL_QND},\cite{Pinard},\cite{Vuletic} seems within the reach of present technology.

I thank L. Lugiato and P. Grangier for useful discussions, M. Guerzoni 
for her contribution to this work and Y. Castin for comments on the 
manuscript. LKB is UMR 8552 du CNRS de l'ENS et de l'UPMC;
support from IFRAF is acknowledged.


\begin{thebibliography}{99}

\bibitem{squeezing} For a recent review on squeezing see e.g. H.~Bachor, ``A Guide
to Experiments in Quantum Optics", Wiley-VCH (2004).
\bibitem{contvar} S.L. Braunstein, P. van Loock,  Rev. of Mod. Phys. {\bf 77},
513 (2005).
\bibitem{QND} P. Grangier, A. Levenson, J.-Ph. Poizat,  Nature {\bf 396}, 537 (1998). 
\bibitem{Kilin}D. B.Horoshko, S.Ya. Kilin, Phys. Rev. A {\bf 61}, 032304 (2000).
\bibitem{PRL_QND}J-F. Roch, K. Vigneron, P. Grelu, A. Sinatra, J-P. 
        Poizat, P. Grangier, Phys. Rev. Lett.{\bf 78}, 634 (1997).
\bibitem{EIT} For a review on EIT see e.g.
M. Fleischhauer, A. Imamoglu, J.P. Marangos,  Rev. Mod. Phys. {\bf 77}, 633 (2005);
\bibitem{Gheri-Marte} K.M.~Gheri, D.F.~Walls, M.A.~Marte, Phys. Rev. A {\bf 50}, 1871 (1994).
\bibitem{Scully} V.A. Sautenkov, Y.V. Rostovstev, M.O. Scully, Phys. Rev. A {\bf 72}, 
065801 (2005), C.L. Garrido Alzar, L.S. Cruz, J.G. Aguirre G\'omez,
	Europhys. Lett. {\bf 61}, 485 (2003).
\bibitem{ProgrOpt} L.A.Lugiato {\it Progess in Optics XXI}, edited by E. Wolf
(North-Holland, Amsterdam, 1977), p. 71.
\bibitem{PRA_QND} A. Sinatra, J-F. Roch, K. Vigneron, P. Grelu, J-P. Poizat, 
        K. Wang, P. Grangier, Phys. Rev. A {\bf 57}, 2980 (1998). 
\bibitem{QNDCoeff} M. Holland, M. Collett, D.F. Walls, M.D. Levenson,
Phys. Rev. A {\bf 42}, 2995 (1990); J.Ph. Poizat, J.-F. Roch and P. Grangier,
Ann. Phys. (Paris) {\bf 19}, 265 (1994).
\bibitem{Gheri} K.M. Gheri, P. Grangier, J.-Ph. Poizat, D. Walls,
Phys. Rev. A {\bf 46}, 4276 (1992).
\bibitem{Nard} L.~Lugiato, P.~Galatola, L.~Narducci, Opt. Comm. {\bf 76}, 276 (1990).
\bibitem{Pinard} V. Josse, A. Dantan, A. Bramati, M. Pinard, E. Giacobino,
Phys. Rev. Lett. {\bf 92}, 123601 (2004);
\bibitem{Vuletic} J.K. Thompson, J. Simon, H. Loh, V. Vuleti\'c, Science {\bf 313}, 74 (2006).


\end{thebibliography}
\end{document}